\documentclass[a4paper,12pt]{iopart}

\usepackage[dvipsnames]{xcolor}

\usepackage{hyperref}
\hypersetup{
  colorlinks=true,
  linkcolor=Red,
  citecolor=Blue,
  urlcolor=Blue,
}

\usepackage{microtype}

\usepackage{iopams}
\usepackage[caption=false,subrefformat=parens]{subfig}

\usepackage{graphicx}

\usepackage[square,numbers,sort&compress]{natbib}
\newcommand*{\eqref}[1]{{(\ref{#1})}}

\usepackage{braket}
\newcommand*{\ketbra}[2]{\ket{#1}\bra{#2}}

\newcommand*{\expect}[1]{\braket{#1}}

\newcommand*{\commutator}[2]{\mathinner{[{#1},{#2}]}}
\newcommand*{\Commutator}[2]{\left[{#1},{#2}\right]}

\newcommand*{\D}[1]{\mathop{d{#1}}\nolimits}

\begin{document}

\title{Open quantum systems with delayed coherent feedback}

\author{S. J. Whalen$^1$, A. L. Grimsmo$^2$ and H. J. Carmichael$^1$}
\address{$^1$ The Dodd-Walls Centre for Photonic and Quantum Technologies, Department of Physics, University of Auckland, Private Bag 92019, Auckland, New Zealand}
\address{$^2$ Institut quantique and D\'{e}partement de Physique, Universit\'{e} de Sherbrooke, Sherbrooke, Qu\'{e}bec J1K 2R1, Canada}
\ead{\href{mailto:simon.whalen@gmail.com}{simon.whalen@gmail.com}, \href{mailto:arne.loehre.grimsmo@usherbrooke.ca}{arne.loehre.grimsmo@usherbrooke.ca} and \href{mailto:h.carmichael@auckland.ac.nz}{h.carmichael@auckland.ac.nz}}

\begin{abstract} 
We present an elementary derivation and generalisation of a recently reported method of simulating feedback in open quantum systems. We use our generalised method to simulate systems with multiple delays, as well as cascaded systems with delayed backscatter. In addition, we derive a generalisation of the quantum regression formula that applies to systems with delayed feedback, and show how to use the formula to compute two-time correlation functions of the system as well as output field properties. Finally, we show that delayed coherent feedback can be simulated as a quantum teleportation protocol that requires only Markovian resources, pre-shared entanglement, and time travel. The requirement for time travel can be avoided by using a probabilistic protocol.
\end{abstract}

\section{Introduction}
\label{sec:introduction}

\begin{figure}
  \centering
  \includegraphics[page=20]{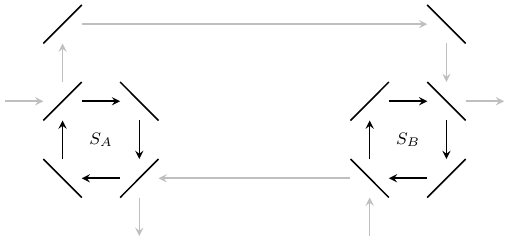}
  \caption{A bipartite open quantum system, depicted here as a pair of ring cavities, with delayed coherent feedback. The propagation delay from subsystem $A$ to subsystem $B$ is greater than that in the reverse direction. Black arrows denote system fields, while grey arrows denote fields propagating in the environment.\label{fig:delayed-coherent-feedback}}
\end{figure}

While the theory of Markovian open quantum systems is well-understood~\cite{carmichael1993,breuer2007}, simulating the dynamics of non-Markovian open quantum systems is considerably more difficult~\cite{de_vega_dynamics_2017}. A signature non-Markovian open quantum system is a qubit emitting into a discrete feedback reservoir: an environment that `remembers' the state of the system and feeds this information back coherently -- as a quantum field -- after one or more discrete time delays. One example of such a system is sketched schematically in Fig.~\ref{fig:delayed-coherent-feedback}. This kind of feedback has previously been studied in work on ``atomic'' emission in front of a mirror~\cite{dorner_laser-driven_2002,carmele_single_2013,tufarelli_dynamics_2013,tufarelli_non-markovianity_2014}, and also in solid state systems with significant propagation delays~\cite{guo_giant_2016}. Such systems are all the more interesting due to the fact that an environmental memory with a continuous kernel, where the evolution of the system depends most generally on its state at all previous times, may be approximated as a sequence of coherent feedback loops with discrete delays. As such, a tool that proves capable of simulating multiple discrete delays may shed light on the more general problem of dealing with a continuous memory.

Discrete propagation delays also appear in the standard theory of cascaded open quantum systems, which describes the situation where the retarded output from one open quantum system drives a second system~\cite{carmichael_quantum_1993,gardiner_driving_1993,kolobov_quantum_1987}. In the standard treatment of cascaded systems there is no backscatter from the second system, so the coupling between the systems is one-way. In this case the propagation delay between the systems is an arbitrary parameter that can be removed by way of a simple transformation of the time variable, which leads to an irreversible, Markovian coupling between subsystems. The cascaded systems formalism is easily generalised to describe irreversible coupling between the systems in both directions, so long as the propagation delay associated with the coupling is sufficiently small that it may be neglected~\cite{carmichael2008se}. We cannot, however, transform away a non-negligible delay in both directions. A treatment of open quantum systems with delayed feedback is therefore necessary to extend the theory of cascaded open quantum systems to encompass cascaded systems with backscatter where there is a nontrivial propagation delay.

There are conditions under which delayed feedback and cascaded systems are well known to be connected. In fact, in a certain limit a coherent feedback loop leads to exactly the same dynamics as a chain of cascaded systems. This equivalence is employed in, for example, work by Menicucci et al.~\cite{menicucci_arbitrarily_2010}. While it is not the case that cascaded systems and delayed feedback are equivalent in general, it turns out that there is sufficient similarity between these two set-ups that results from the theory of cascaded systems can be exploited to help simulate feedback. Recently, Grimsmo~\cite{grimsmo_time-delayed_2015} employed tensor network methods to show how to simulate a nonlinear quantum system (such as a two-state `atom') interacting with a discrete feedback reservoir and indeed, the resulting equations demonstrate a close connection to the master equation for cascaded systems. Pichler and Zoller~\cite{pichler_photonic_2016} subsequently published a different technique, based on matrix product states, for simulating quantum circuits in the regime where time delays are significant. We present here an elementary derivation and generalisation of Grimsmo's method, which permits the simulation of multiple delays, as well as cascaded systems with delayed backscatter where the delay may differ in each direction.

Our derivation is based on the fact that the evolution of a \emph{generic} open quantum system may be decomposed into a nested sequence of evolutions over distinct time intervals---what we refer to as a decomposition into intervals. This decomposition is outlined in Sec.~\ref{sec:intervals}. In Sec.~\ref{sec:feedback}, we apply this decomposition to open quantum systems interacting with the environment such that the system experiences delayed coherent feedback, with discrete delays of various lengths applying to interactions between different pairs of subsystems. We go on in Sec.~\ref{sec:examples} to present a series of examples of the use of this algorithm, and demonstrate in Sec.~\ref{sec:two-time-correlation} how the algorithm may be extended to enable the calculation of two-time (and more generally multi-time) correlation functions, presenting an example calculation of the second-order photon correlation for the output field of a system interacting with a delayed coherent feedback loop. We then show in Sec.~\ref{sec:teleportation} that our algorithm for simulating open quantum systems with delayed coherent feedback may be interpreted as a quantum teleportation protocol. We summarise our main results and discuss their application to more general systems in Sec.~\ref{sec:conclusion}.

\section{Decomposing the evolution of an open quantum system into intervals}
\label{sec:intervals}

The evolution of an open quantum system may be decomposed into intervals. This decomposition, based on the work of Grimsmo~\cite{grimsmo_time-delayed_2015}, is outlined below. In this section we will justify it for a generic system, before applying the resulting simulation algorithm to the case of a system with delayed coherent feedback in Sec.~\ref{sec:feedback}.

\begin{figure}
  \centering
  \includegraphics[page=21]{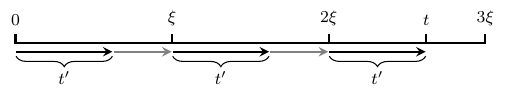}
  \caption{Decomposition of an open quantum system's evolution into intervals. Here, we split the evolution of a system up to time $t$ into $n=3$ intervals of length $\xi$, making use of the auxiliary time variable $t' = t - (n - 1)\xi$. Black arrows denote the evolution of all system copies up to time $t'$, corresponding to the map $\Phi^{(n)}(t', 0)$. Gray arrows illustrate the evolution of all systems save the last from $t'$ up to $\xi$, corresponding to $\Phi^{(n-1)}(\xi, t')$.\label{fig:t-prime}}
\end{figure}

We begin by summarising the structure of the decomposition implemented as dynamical map. Following Gardiner
and Zoller~\cite{gardiner2004} we introduce a complete set, $\{e_{j}\}$, of basis operators for the system that are orthogonal with respect to the trace: $\tr e_{j}^\dagger e_{k} = \delta_{jk}$. The state of the reduced system (with the environment traced out) can be expanded, as usual, in terms of these basis operators:
\begin{equation}
  \label{eq:1}
  \rho(t) = \sum_{j_1} \tr\left[\rho(0) e_{j_1}^\dagger\right] e_{j_1}(t) \,.
\end{equation}
The evolution of the basis operators (and hence the system) is divided into intervals of length $\xi$, with each interval represented by a formally separate Hilbert space. To divide the system's evolution into $n \equiv \lceil t / \xi \rceil$ intervals, we require $n$ copies of the system's Hilbert space. The total evolution time $t$ will not necessarily be an integer multiple of the interval length, so we define the auxiliary time variable $t' \equiv t - (n - 1)\xi$, as illustrated in Fig.~\ref{fig:t-prime}.

The evolved basis operator $e_{j_1}(t)$ is obtained using the mapping formula
\begin{equation}
  \label{eq:2}
  e_{j_1}(t) = \tr_1 \cdots \tr_{n-1} \sum_{j_2 \cdots j_n} e_{j_1 \cdots j_n}(\xi; t') [e_{j_2 \cdots j_n}^\dagger \otimes I] \,,
\end{equation}
where we have defined the product basis operators $e_{j_1 \cdots j_n} \equiv e_{j_1} \otimes \cdots \otimes e_{j_n}$, and where $e_{j_1 \cdots j_n}(\xi; t')$ is a basis operator of the entire fictitious `chain' of system copies, given by
\begin{equation}
  \label{eq:3}
  e_{j_1 \cdots j_n}(\xi; t') = \Phi^{(n-1)}(\xi, t') \Phi^{(n)}(t', 0) e_{j_1 \cdots j_n} \,.
\end{equation}
At this stage, we simply assume the existence of the dynamical map $\Phi^{(m)}(t_1, t_0)$ that describes the evolution of the first $m$ system copies in the chain from $t_0$ up to $t_1$; in Sec.~\ref{sec:effective-dissipation-kernel} we will demonstrate a convenient method of deriving this map, by using an effective representation of the environment's entire timeline. We can see that Eq.~\eqref{eq:3} describes the same decomposition of the system's timeline as Fig.~\eqref{fig:t-prime}: the dynamical map is used to evolve the chain of system copies in its entirety up to time $t'$, after which all systems up to but not including the last in the chain (the `present' interval) are evolved all the way up to time $\xi$. The complete timeline of each basis operator's evolution is reconstructed using Eq.~\eqref{eq:2}, and these basis operators are then used in Eq.~\eqref{eq:1} to obtain the state of the system. The map $\Phi^{(m)}(t_1, t_0)$ is met in its simplest form by considering a closed quantum system and starting out with more familiar notation.

\subsection{A closed quantum system}
\label{sec:closed-decomposition}

We consider the evolution of a density operator, $\rho(t)$, with initial condition written in matrix notation,
\begin{equation}
  \label{eq:4}
  \rho(0) = \sum_{\mu\nu} \braket{\mu|\rho(0)|\nu} \left(\ketbra{\mu}{\nu}\right) \,,
\end{equation}
where $\braket{\mu|\nu} = \delta_{\mu\nu}$. The unitary time-evolution operator is denoted $U(t, t_0) \equiv T \exp[ -i \int_{t_0}^t \D{s} H(s) ]$, where the time ordering operator $T$ orders products of time dependent operators such that their time arguments increase from right to left. We break the evolution from the initial time $0$ to time $t$ into two intervals: an initial interval of length $\xi$ and the remainder. The time evolution is generated by propagating the elementary operators $\left(\ketbra{\mu}{\nu}\right)$. By interrupting the evolution at $\xi$ and re-expanding the result in matrix notation before proceeding, we get
\begin{eqnarray}
  \label{eq:5}
  \fl U(t,0)\left(\ketbra{\mu}{\nu}\right)U^\dagger(t,0) \nonumber \\
  = U(t,\xi) \left[ U(\xi,0)\left(\ketbra{\mu}{\nu}\right)U^\dagger(\xi,0) \right] U^\dagger(t,\xi) \nonumber \\
  = U(t,\xi) \Bigl[ \textstyle\sum_{\mu'\nu'} \braket{\mu'|U(\xi,0)\left(\ketbra{\mu}{\nu}\right)U^\dagger(\xi,0)|\nu'} \ketbra{\mu'}{\nu'} \Bigr] U^\dagger(t,\xi) \nonumber \\
  = \sum_{\mu'\nu'} U(t,\xi) \left(\ketbra{\mu'}{\nu'}\right) U^\dagger(t,\xi) \tr\left[U(\xi,0)\left(\ketbra{\mu}{\nu}\right)U^\dagger(\xi,0) {\left( \ketbra{\mu'}{\nu'} \right)}^\dagger\right] \,.
\end{eqnarray}
Then returning to our more compact notation, we may write
\begin{equation}
  \label{eq:6}
  U(t,0) e_{j_1} U^\dagger(t,0) = \sum_{j_2} U(t,\xi) e_{j_2} U^\dagger(t,\xi) \tr\left[ U(\xi,0) e_{j_1} U^\dagger(\xi,0) e_{j_2}^\dagger \right] \,,
\end{equation}
where we allow a single index to stand in for the pair $(\mu,\nu)$, with $e_{j_1}$ -- a basis operator as introduced above -- standing in for $\left(\ketbra{\mu}{\nu}\right)$ and $e_{j_2}$ for $\left(\ketbra{\mu^\prime}{\nu^\prime}\right)$. Finally, we note that the results required from the two separate steps in the time evolution may be computed in formally distinct Hilbert spaces, which we index as space 1 (interval 0 to $\xi$) and space 2 (interval $\xi$ to $t$), and then ``parked'' in appropriate locations in a tensor product of those two spaces. This allows us to re-express Eq.~\eqref{eq:6} using tensor product notation:
\begin{equation}
  \label{eq:7}
  e_{j_1}(t) = \tr_1 \sum_{j_2} \left[ \left( U(\xi,0) e_{j_1} U^\dagger(\xi,0) \right) \otimes \left( U(t,\xi) e_{j_2} U^\dagger(t,\xi) \right) \right] \bigl[ e_{j_2}^\dagger\otimes I \bigr] \,.
\end{equation}
The division into $n$ intervals gives an obvious extension of the formula, which turns out to be exactly Eq.~\eqref{eq:2} with 
\begin{eqnarray}
  \label{eq:8}
  \fl e_{j_1 \cdots j_n}(\xi; t') = \left[ U(\xi,0) \otimes U(2\xi,\xi) \otimes \cdots \otimes U(t, (n-1)\xi) \right] e_{j_1 \cdots j_n} \nonumber \\ \times \left[ U^\dagger(\xi,0) \otimes U^\dagger(2\xi,\xi) \otimes \cdots \otimes U^\dagger(t,(n-1)\xi) \right] \,,
\end{eqnarray}
where we have used the fact that operators of different system copies commute to group all the unitary operators together. Equation~\eqref{eq:8} evolves the basis operators for each of the system copies $1$ to $(n-1)$ through an interval of length $\xi$, and Eq.~\eqref{eq:2} maps these final conditions onto the corresponding initial conditions for system copies $2$ through $n$, respectively. The $n$\textsuperscript{th} system is evolved through an interval of length $t'$, where we recall that $t \equiv t' + (n - 1)\xi$. As such, this process reconstructs the evolved basis operator $e_{j_1}(t)$ and (by completeness) the state of the `real' system.

In Eq.~\eqref{eq:8}, we have separate unitaries describing the evolution of each system copy. In Sec.~\ref{sec:effective-dissipation-kernel}, however, we will consider open systems with delayed feedback, with the feedback simulated by way of a single effective environment shared between all system copies in the chain. The feedback will appear in this formalism as interactions between different system copies, meaning that we will no longer be able to treat the evolution of each interval independently. For that reason we now generalise our notation, creating a combined unitary operator that evolves all systems in the chain simultaneously. To write this generalisation in a compact way, we first need to define some notation which will be used throughout the rest of this paper. We define
\begin{equation}
  \label{eq:9}
  A_{m} \equiv {\underbrace{I \otimes \cdots \otimes I}_{(m-1) \; \mathrm{times}}} \otimes A \otimes {\underbrace{I \otimes \cdots \otimes I}_{(n - m) \; \mathrm{times}}} \,,
\end{equation}
for any system operator $A$, where $I$ is the identity operator for a single system copy. We now define an evolution operator acting on all $n$ system copies:
\begin{equation}
  \label{eq:10}
  U^{(m)}(t_1, t_0) \equiv T \exp\left[ -i \int_{t_0}^{t_1} \D{s} \sum_{l=1}^{m} H_{l}(s + (l - 1) \xi) \right] \,,
\end{equation}
where in this context $0 \leq t_0 \leq t_1 \leq \xi$, as this unitary will only be used to evolve the chain on the interval $[0, \xi]$. It is easily verified that
\begin{equation}
  \label{eq:11}
  U^{(n-1)}(\xi, t') U^{(n)}(t', 0) = U(\xi,0) \otimes U(2\xi,\xi) \otimes \cdots \otimes U(t, (n-1)\xi) \,,
\end{equation}
which means that the evolved basis operators in Eq.~\eqref{eq:8} can now be written in the form
\begin{equation}
  \label{eq:12}
  e_{j_1 \cdots j_n}(\xi; t') = U^{(n-1)}(\xi, t') U^{(n)}(t', 0) e_{j_1 \cdots j_n} U^{(n)\dagger}(t', 0) U^{(n-1)\dagger}(\xi, t') \,.
\end{equation}
It follows that in the simple case of a closed system the map introduced in Eq.~\eqref{eq:3} is given by $\Phi^{(m)}(t_1, t_0) e_{j_1 \cdots j_n} = U^{(m)}(t_1, t_0) e_{j_1 \cdots j_n} U^{(m)\dagger}(t_1, t_0)$.

\subsection{An open quantum system}
\label{sec:open-decomposition}

We now begin the process of generalising to open quantum systems. Firstly we define, in general terms, the system whose evolution we wish to decompose. Suppose the evolution of an open quantum system is described, in a rotating frame, by the Hamiltonian $H(t) = H_S + H_{SE}(t)$, where $H_S$ involves only system operators and describes the internal dynamics of the system, and the interaction between system and environment is generated by
\begin{equation}
  \label{eq:13}
  H_{SE}(t) = \sum_\alpha \left[ a_\alpha^\dagger B_\alpha(t) + B_\alpha^\dagger(t) a_\alpha \right] \,,
\end{equation}
where $B_\alpha(t) = \sum_j \kappa_{\alpha j} e^{i (\omega_0 - \omega_j) t} b_j$, with couplings $\kappa_{\alpha j}$. Here $\omega_0$ is some fiducial frequency that may be freely chosen. The index $\alpha$ labels different subsystems, while $j$ labels modes of the environment. System and environment operators commute at equal times, and the environment will be assumed to be an assemblage of harmonic oscillators with $\commutator{b_j}{b_j^\dagger} = \delta_{jj'}$. It is also useful to define the dissipation (memory) kernel of the reservoir: $F_{\alpha\beta}(t_2 - t_1) \equiv \commutator{B_\alpha(t_2)}{B_\beta^\dagger(t_1)}$. We assume that the initial combined state of the system and environment is separable and given by $\rho(0) = \rho_E \otimes \rho_S(0)$, where $\rho_S(0)$ and $\rho_E$ are the initial states of the system and environment respectively.

The derivation for an open quantum system proceeds much as in Sec.~\ref{sec:closed-decomposition}. We start by making the trivial expansion
\begin{equation}
  \label{eq:14}
  \rho(0) = \sum_{\mu\nu} \braket{\mu|\rho_S(0)|\nu} \left( \rho_E \otimes \ketbra{\mu}{\nu} \right) \,,
\end{equation}
where in this case the operators $(\ketbra{\mu}{\nu})$ form a basis for the system. We once again break the evolution from $0$ to $t$ into two intervals, interrupting the evolution at $\xi$ and re-expanding the system---but not the environment:
\begin{eqnarray}
  \label{eq:15}
  \fl U(t,0)\left(\rho_E \otimes \ketbra{\mu}{\nu}\right)U^\dagger(t,0) \nonumber \\
  = U(t,\xi) \left\{ U(\xi,0) \left( \rho_E \otimes \ketbra{\mu}{\nu} \right) U^\dagger(\xi,0) \right\} U^\dagger(t,\xi) \nonumber \\
  = \sum_{\mu'\nu'} U(t,\xi) \left\{ \tr_S \Bigl[ U(\xi,0) \left( \rho_E \otimes \ketbra{\mu}{\nu} \right) U^\dagger(\xi,0) {(\ketbra{\mu'}{\nu'})}^\dagger \Bigr] \ketbra{\mu'}{\nu'} \right\} \nonumber \\ \qquad \times U^\dagger(t,\xi) \,.
\end{eqnarray}
Just as we did above in the case of the open quantum system, we replace the pair of indices $(\mu, \nu)$ with a single index by introducing basis operators $\{e_{S;j}\}$ for the system, which gives
\begin{eqnarray}
  \label{eq:16}
  \fl U(t,0) (\rho_E \otimes e_{S;j_1}) U^\dagger(t,0) \nonumber\\ = \sum_{j_2} U(t,\xi) \left\{ \tr_S \Bigl[ U(\xi,0) (\rho_E \otimes e_{S;j_1}) U^\dagger(\xi,0) e_{S;j_2}^\dagger  \Bigr] e_{S;j_2} \right\} U^\dagger(t,\xi) \,.
\end{eqnarray}
We now introduce a new unitary operator
\begin{eqnarray}
  \label{eq:17}
  \fl U'(t_1, t_0) \equiv T \exp\Biggl( -i \int_{t_0}^{t_1} \D{s} \sum_m 1_{(m-1) \xi \leq s < m \xi} \nonumber\\ \hspace*{3cm} \times \Biggl\{H_{S;m} + \sum_\alpha \left[ B_\alpha(s) \otimes a_{\alpha;m}^\dagger + B_\alpha^\dagger(s) \otimes a_{\alpha;m} \right]\Biggr\} \Biggr) \,.
\end{eqnarray}
where $a_{\alpha;m}$ is an operator of the $m$\textsuperscript{th} system copy, expressed using the notation introduced in Eq.~\eqref{eq:9}, and where $1_{(m-1) \xi \leq s < m \xi}$ is an indicator function, defined as $1$ when $(m-1) \xi \leq s < m \xi$ and $0$ otherwise.  This unitary allows us to re-write Eq.~\eqref{eq:16} using tensor products. Generalising at the same time to $n$ intervals, we find
\begin{eqnarray}
  \label{eq:18}
  \fl U(t,0) (\rho_E \otimes e_{S;j_1}) U^\dagger(t,0) = \nonumber\\ \tr_{S;1} \cdots \tr_{S;n-1} \sum_{j_2 \cdots j_n} \left[ U'(t, 0) (\rho_E \otimes e_{S;j_1 \cdots j_n}) U^{\prime\dagger}(t, 0) \right] [I_E \otimes e_{S;j_2 \cdots j_n}^\dagger \otimes I_S] \,. \nonumber\\
\end{eqnarray}
Finally, we trace out the environment to obtain the state of the reduced system, $\rho_S(t) \equiv \tr_E \rho(t)$:
\begin{equation}
  \label{eq:19}
  \rho_S(t) = \sum_{j_1} \tr_S \left[ \rho(0) e_{S;j_1}^\dagger \right] e_{S; j_1}(t) \,,
\end{equation}
where
\begin{equation}
  \label{eq:20}
  e_{S;j_1}(t) = \tr_{S;1} \cdots \tr_{S;n-1} \sum_{j_2 \cdots j_n} e_{S;j_1 \cdots j_n}(\xi; t') [e_{S;j_2 \cdots j_n}^\dagger \otimes I_S] \,,
\end{equation}
and where, recalling that $t \equiv t' + (n-1)\xi$, we have defined
\begin{equation}
  \label{eq:21}
  \fl e_{S;j_1 \cdots j_n}(\xi; t') \equiv \tr_E \left[ U'(t' + (n-1)\xi, 0) (\rho_E \otimes e_{S;j_1 \cdots j_n}) U^{\prime\dagger}(t' + (n-1)\xi, 0) \right] \,.
\end{equation}
Equations~\eqref{eq:19} and~\eqref{eq:20} are, up to minor notational differences, identical to Eqs.~\eqref{eq:1} and~\eqref{eq:2}. In Sec.~\ref{sec:effective-dissipation-kernel} we will derive an effective representation of the system--reservoir dynamics that will allow us to re-write Eq.~\eqref{eq:21} in terms of a dynamical map, in the form of Eq.~\eqref{eq:3}.

\subsection{Effective dissipation kernel for an open quantum system}
\label{sec:effective-dissipation-kernel}

In Sec.~\ref{sec:open-decomposition} we demonstrated a decomposition of the evolution of an open quantum system into intervals, but we have not yet derived a map describing the evolution of the reduced system. We now introduce a new system--environment interaction that will allow us to derive a reduced system map. We refer to this as an \emph{effective} system--environment interaction, because it will be defined in such a way that its effects on the system reproduce the evolution of the real system once the environment has been traced out. Provided that the resulting reduced system map turns out to be divisible, the algorithm described in Sec.~\ref{sec:closed-decomposition} may be extended to open quantum systems. As we will see in Sec.~\ref{sec:feedback}, when dealing with delayed coherent feedback this effective interaction leads to a considerable simplification, because it means the feedback may be modelled as an interaction between intervals/system copies.

To find this effective system--environment interaction we define a new Hamiltonian $\tilde{H}_{m}(t') \equiv H_{S;m} + \tilde{H}_{SE;m}(t')$ for a single system copy with interaction term
\begin{equation}
  \label{eq:22}
  \tilde{H}_{SE;m}(t') \equiv \sum_\alpha \bigl[ \tilde{B}_{\alpha;m}(t') \otimes a_{\alpha;m}^\dagger + {\tilde{B}_{\alpha;m}^\dagger}(t') \otimes a_{\alpha;m} \bigr] \,.
\end{equation}
We also define the corresponding unitary evolution operator
\begin{equation}
  \label{eq:23}
  \tilde{U}_m(t_1, t_0) \equiv T \exp\left[ -i \int_{t_0}^{t_1} \D{s} \tilde{H}_m(s) \right] \,.
\end{equation}
Here, `effective' quantities that differ from their analogues from Sec.~\ref{sec:open-decomposition} are denoted by tildes. Note that, consistent with the notation introduced in Eq.~\eqref{eq:9}, the Hamiltonian~\eqref{eq:22} acts only on the $m$\textsuperscript{th} system copy. In an analogue of Eq.~\eqref{eq:10}, we define a corresponding effective unitary for the chain of fictitious system copies,
\begin{equation}
  \label{eq:24}
  \tilde{U}^{(m)}(t_1, t_0) \equiv T \exp\left[ -i \int_{t_0}^{t_1} \D{s} \sum_{l=1}^{m} \tilde{H}_{l}(s) \right] \,.
\end{equation}
Equation~\eqref{eq:24} is similar to Eq.~\eqref{eq:10}; however, note the difference in the time argument of the Hamiltonian under the integral. Using this unitary, we define a reduced system map:
\begin{equation}
  \label{eq:25}
  \Phi^{(m)}(t_1, t_0) \chi \equiv \tr_E \left\{\tilde{U}^{(m)}(t_1, t_0) [ \rho_{\tilde{E}} \otimes \chi] \tilde{U}^{(m)\dagger}(t_1, t_0)\right\} \,,
\end{equation}
where $\chi$ is some operator of the reduced system copies, and $\rho_{\tilde{E}}$ is the initial state of the effective environment. We require that this reduced system map reproduce Eq.~\eqref{eq:21} when used in Eq.~\eqref{eq:3}, in the sense that we should have
\begin{eqnarray}
  \label{eq:26}
  \fl \Phi^{(n-1)}(\xi, t') \Phi^{(n)}(t', 0) e_{S;j_1 \cdots j_n} \nonumber\\ = \tr_E \left[ \tilde{U}^{(n-1)}(\xi, t') \tilde{U}^{(n)}(t', 0) \left( \rho_{\tilde{E}} \otimes e_{S;j_1 \cdots j_n} \right) \tilde{U}^{(n)\dagger}(t', 0) \tilde{U}^{(n-1)\dagger}(\xi, t') \right] \,, \nonumber\\
\end{eqnarray}
as well as
\begin{equation}
  \label{eq:27}
  \fl e_{S;j_1 \cdots j_n}(\xi; t') = \tr_E \left[ \tilde{U}^{(n-1)}(\xi, t') \tilde{U}^{(n)}(t', 0) \left( \rho_{\tilde{E}} \otimes e_{S;j_1 \cdots j_n} \right) \tilde{U}^{(n)\dagger}(t', 0) \tilde{U}^{(n-1)\dagger}(\xi, t') \right] \,,
\end{equation}
with $e_{S;j_1 \cdots j_n}(\xi; t')$ given by Eq.~\eqref{eq:21}. As discussed in~\ref{sec:deriv-divisible}, Eq.~\eqref{eq:26} is satisfied whenever the map $\Phi^{(m)}(t_1, t_0)$ is \emph{divisible}~\cite{rivas_quantum_2014,wolf_dividing_2008} for all $m$, meaning that
\begin{equation}
  \label{eq:28}
  \Phi^{(m)}(t_2, t_0) = \Phi^{(m)}(t_2, t_1) \Phi^{(m)}(t_1, t_0) \,, \qquad t_2 \geq t_1 \geq t_0 \,.
\end{equation}
In the following, we will show that the effective environment may be structured such that Eq.~\eqref{eq:27} also holds.

The first restriction we impose on the effective environment operators $\tilde{B}_{\alpha;m}(t')$ is the stipulation that
\begin{equation}
  \label{eq:29}
  \Commutator{\tilde{B}_{\alpha;m}(t_2)}{\tilde{B}_{\beta;m'}^\dagger(t_1)} = 0 \qquad \mathrm{when} \quad t_2 \geq t_1 \quad \mathrm{and} \quad m < m' \,.
\end{equation}
From this commutation relation, we easily find that a similar relation holds for the effective Hamiltonian:
\begin{equation}
  \label{eq:30}
  \Commutator{\tilde{H}_{m}(t_2)}{\tilde{H}_{m'}(t_1)} = 0 \qquad \mathrm{when} \quad t_2 \geq t_1 \quad \mathrm{and} \quad m < m' \,.
\end{equation}
Equation~\eqref{eq:29} thus means that a system copy appearing earlier in the chain of decomposed intervals is independent of (commutes with) any subsequent system copy in the chain when the former is considered at a later time than the latter. This ensures that a given system copy cannot be affected by any system copy in its relative `future' -- a natural requirement of causality for the real system. Put another way, causality in the chain of system copies runs from $m=1$ towards $m=n$.

As such, it follows from Eq.~\eqref{eq:30} that, analogous to Eq.~\eqref{eq:11}, we can write
\begin{equation}
  \label{eq:31}
  \tilde{U}^{(n-1)}(\xi, t') \tilde{U}^{(n)}(t', 0) = \tilde{U}_n(t', 0) \tilde{U}_{n-1}(\xi, 0) \cdots \tilde{U}_2(\xi, 0) \tilde{U}_1(\xi, 0) \,,
\end{equation}
which allows us to re-write Eq.~\eqref{eq:27} as
\begin{equation}
  \label{eq:32}
  \fl e_{S;j_1 \cdots j_n}(\xi; t') = \tr_E \left\{ \tilde{U}_n(t', 0) \cdots \tilde{U}_1(\xi, 0) \left[ \rho_{\tilde{E}} \otimes e_{S;j_1 \cdots j_n} \right] \tilde{U}_1^\dagger(\xi, 0) \cdots \tilde{U}_n^\dagger(t', 0) \right\} \,.
\end{equation}
This equation is the open systems version of Eq.~\eqref{eq:8}. It remains to choose the environment operator $\tilde{B}_{\alpha;m}(t')$ such that Eq.~\eqref{eq:32} does indeed reproduce Eq.~\eqref{eq:21}. We do not have to specify this operator explicitly: if we restrict attention to vacuum reservoirs -- by assuming that both $\rho_E$ and $\rho_{\tilde{E}}$ represent the vacuum states of the respective reservoirs -- the interaction of the system copies with the environment may be fully characterised by the dissipation kernel. As such, our task becomes to derive the effective dissipation kernel $\tilde{F}_{\alpha\beta;mm'}(t_2 - t_1) \equiv \commutator{\tilde{B}_{\alpha;m}(t_2)}{\tilde{B}_{\beta;m'}^\dagger(t_1)}$ such that Eqs.~\eqref{eq:21} and Eq.~\eqref{eq:32} are equivalent. After a few simple manipulations of these two equations, as detailed in~\ref{sec:derivation-eff-kernel}, we find that this condition is satisfied by the effective dissipation kernel
\begin{equation}
  \label{eq:33}
  \tilde{F}_{\alpha\beta;mm'}(t_2 - t_1) \equiv
  \cases{
    F_{\alpha\beta}(t_2 - t_1 + (m - m')\xi) & $m \geq m'$ \,, \\
    0 & otherwise,
  }
\end{equation}
with $t_2 \geq t_1$, where the case $m < m'$ follows from Eq.~\eqref{eq:29}. Note that the dissipation kernel is, by construction, a Hermitian function: $\tilde{F}_{\alpha\beta;mm'}(t_1 - t_2) = \tilde{F}_{\beta\alpha;m'm}(t_2 - t_1)$. In Eq.~\eqref{eq:33} we focus on the case $t_2 \geq t_1$, as this is the only case relevant to the dynamics. Henceforth we will work entirely with the effective dissipation kernel~\eqref{eq:33}, and omit the tildes denoting effective quantities for the sake of simplicity.

We have thus shown that Eq.~\eqref{eq:27} is satisfied when the interaction of the system copies with the effective environment is described by the dissipation kernel~\eqref{eq:33}. If the resulting map, given by Eq.~\eqref{eq:25}, is divisible then Eq.~\eqref{eq:26} is also satisfied. Combining Eqs.~\eqref{eq:26} and~\eqref{eq:27}, we obtain
\begin{equation}
  \label{eq:34}
  \Phi^{(n-1)}(\xi, t') \Phi^{(n)}(t', 0) e_{S;j_1 \cdots j_n} = e_{S;j_1 \cdots j_n}(\xi; t') \,,
\end{equation}
where $e_{S;j_1 \cdots j_n}(\xi; t')$ is given by Eq.~\eqref{eq:21}. Eq.~\eqref{eq:34} is, up to minor notational differences, Eq.~\eqref{eq:3}.

\section{Delayed coherent feedback}
\label{sec:feedback}

We want to use the decomposition into intervals described in the previous section to simulate a coherent environmental feedback loop with discrete delays. With that in mind we specify the dissipation kernel
\begin{equation}
  \label{eq:35}
  F_{\alpha\beta}(t') \equiv \sum_j \bigl[ \gamma_{\alpha\beta j} \delta(t' - \tau_{\alpha\beta j}) + \gamma_{\alpha\beta j}^* \delta(t' + \tau_{\alpha\beta j}) \bigr] \,,
\end{equation}
with $\tau_{\alpha\beta j} \geq 0$, which describes exactly such a system. We assume that the delays $\tau_{\alpha\beta j}$ are commensurable, with the interval length $\xi$ chosen to match the greatest common divisor of the delays so that $k_{\alpha\beta j} = \tau_{\alpha\beta j} / \xi$ is an integer for all $\alpha$, $\beta$, and $j$.

Our task now is to derive the map $\Phi^{(m)}(t', t_0)$, $t_0 \leq t' \leq \xi$, describing the reduced system dynamics associated with the dissipation kernel Eq.~\eqref{eq:35}. Substituting the dissipation kernel into Eq.~\eqref{eq:33} yields
\begin{eqnarray}
  \label{eq:36}
  F_{\alpha\beta;mm'}(t') =
  \cases{
    \sum_j \bigl[ \gamma_{\alpha\beta j} \delta(t' + (m - m' - k_{\alpha\beta j})\xi) \\ \qquad + \gamma_{\alpha\beta j}^* \delta(t' + (m - m' + k_{\alpha\beta j})\xi) \bigr] & $m \geq m'$ \,, \\
    0 & otherwise.
  }
\end{eqnarray}
Recalling that, in the context of Eq.~\eqref{eq:3}, $t' \leq \xi$, and noting that both $m - m'$ and $k_{\alpha\beta j}$ are non-negative integers, we can see that only those terms in Eq.~\eqref{eq:36} for which $k_{\alpha\beta j} = m - m'$ contribute to the evolution of system operators on each interval. As such, it is easily seen that the dissipation kernel~\eqref{eq:36} describes $n$ identical systems coupled to a common reservoir in which the output of each propagates in the direction of increasing $m$.

As is well-known, this situation is described by the theory of cascaded open quantum systems. It is therefore straightforward to show that, when the initial state of the combined system is separable and the environment is in its vacuum state, Eq.~\eqref{eq:22} leads to the following Liouvillian generator of the dynamics:
\begin{eqnarray}
  \label{eq:37}
  \fl \mathcal{L}^{(m)} \chi = \sum_{l=1}^m \biggl( -i \commutator{H_{S;l}}{\chi} + \sum_{\alpha\beta j} 1_{l-k_{\alpha\beta j} \geq 1} \Bigl\{ \gamma_{\alpha\beta j} \commutator{a_{\beta;l-k_{\alpha\beta j}} \chi}{a_{\alpha;l}^\dagger} \nonumber \\ + \gamma_{\alpha\beta j}^* \commutator{a_{\alpha;l}}{\chi a_{\beta;l-k_{\alpha\beta j}}^\dagger} \Bigr\} \biggr) \,.
\end{eqnarray}
The correspondence shown here between delayed feedback and cascaded systems is analogous to the well-known method of solving a classical delay-differential equation by re-casting it as a multivariate Markov process~\cite{frank_multivariate_2002}. The fact that the dissipation kernel~\eqref{eq:35} describing delayed coherent feedback maps to the generator~\eqref{eq:37} for cascaded open quantum systems is what makes the algorithm presented in Sec.~\ref{sec:intervals} so useful in this particular case: we already know how to solve the dynamics obtained by splitting the evolution into intervals.

Having found the Liouvillian~\eqref{eq:37}, we are in a position to write down the map $\Phi^{(m)}(t', t_0)$ that first appeared in Eq.~\eqref{eq:25}:
\begin{equation}
  \label{eq:38}
  \Phi^{(m)}(t', t_0) = \exp\left[ \mathcal{L}^{(m)} (t' - t_0) \right] \,.
\end{equation}
The density matrix of the real system is then obtained by substituting this map into Eq.~\eqref{eq:3}, and subsequently using Eqs.~\eqref{eq:2} and~\eqref{eq:1}. The Liouvillian generator~\eqref{eq:37} may be written in Lindblad form, and as such the map~\eqref{eq:38} is divisible~\cite{rivas_quantum_2014,chruscinski_markovianity_2012} and, as discussed in \ref{sec:effective-dissipation-kernel}, satisfies Eq.~\eqref{eq:3}.

In the case of a single system with a single delay, this algorithm is equivalent to that derived by Grimsmo~\cite{grimsmo_time-delayed_2015}. The presentation given above is, however, more general, in the sense that it admits multiple subsystems and multiple delays. We show below how to use this formalism to describe feedback with more than one delay, as well as cascaded systems with delayed backscatter in the case where the delay differs in either direction.

The algorithm reported here is not computationally efficient for long times. Because a copy of the system Hilbert space is required for each and every $\xi$-interval we wish to simulate, both the number of basis operators that must be evolved and the dimension of these operators increases exponentially in the number of intervals. This means that it is not practical to simulate beyond a few $\xi$. In particular, while we have established that the algorithm can handle multiple commensurable delays, the exponential scaling of memory requirements will pose difficulties if any individual $k_{\alpha\beta j}$ is large. For this reason, in the following examples we have restricted ourselves to situations where all $k_{\alpha \beta j}$ are small integers. Furthermore, the restriction to small numbers of intervals means that in many situations the steady state is not accessible using this algorithm. The advantage of our presentation is that it highlights the connection between networks of cascaded systems and delayed coherent feedback, generalising the earlier work of Grimsmo. Although the Liouvillian~\eqref{eq:37} is exactly that of an array of cascaded systems, delayed feedback is not equivalent to such an array at the level of physical systems. The mapping rule~\eqref{eq:2}, enabled by the decomposition into basis operators, is required to correctly account for correlations between the different time intervals when these intervals are represented as separate systems; it does so simply by mapping the final state of each fictitious system copy onto the initial state of the next system copy in the chain. A completely analogous algorithm can be used to simulate classical delay-differential equations.

\section{Examples}
\label{sec:examples}

\begin{figure}[p]
  \begin{minipage}[b]{.49\textwidth}
  \centering
  \subfloat[][\label{fig:delay}]{\includegraphics[page=1]{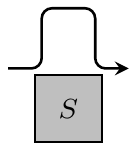}}
  \subfloat[][\label{fig:delay cascade}]{\includegraphics[page=2]{figures}}\\
  \subfloat[][\label{fig:delay results}]{\includegraphics[page=3]{figures}}
  \end{minipage}
  \begin{minipage}[b]{.49\textwidth}
  \centering
  \subfloat[][\label{fig:2delay}]{\includegraphics[page=4]{figures}}
  \subfloat[][\label{fig:2delay cascade}]{\includegraphics[page=5]{figures}}\\
  \subfloat[][\label{fig:2delay results}]{\includegraphics[page=6]{figures}}
  \end{minipage}\\
  \begin{minipage}[b]{.49\textwidth}
  \centering
  \subfloat[][\label{fig:3delay}]{\includegraphics[page=7]{figures}}
  \subfloat[][\label{fig:3delay cascade}]{\includegraphics[page=8]{figures}}\\
  \subfloat[][\label{fig:3delay results}]{\includegraphics[page=9]{figures}}
  \end{minipage}
  \begin{minipage}[b]{.49\textwidth}
  \centering
  \subfloat[][\label{fig:loop}]{\includegraphics[page=10]{figures}}
  \subfloat[][\label{fig:loop cascade}]{\includegraphics[page=11]{figures}}\\
  \subfloat[][\label{fig:loop results}]{\includegraphics[page=12]{figures}}
  \end{minipage}
  \caption{Qubits with delayed coherent feedback. All simulations have parameters $\gamma\tau = 5$ and $\phi = \pi$. \protect\subref{fig:delay}--\protect\subref{fig:delay results} Single feedback loop. \protect\subref{fig:2delay}--\protect\subref{fig:2delay results} Two feedback loops, with round trip times $\tau$ and $2\tau$. \protect\subref{fig:3delay}--\protect\subref{fig:3delay results} Three feedback loops, with round-trip times $\tau$, $2\tau$, and $3\tau$. \protect\subref{fig:loop}--\protect\subref{fig:loop results} An ``infinite loop'', as described by the dissipation kernel~\eqref{eq:40} in the limit $N \to \infty$, where the system is coupled to a multi-mode cavity, which acts as a one-dimensional waveguide that feeds the system output back after one round trip, then after two round trips, three, and so on. \protect\subref{fig:delay}, \protect\subref{fig:2delay}, \protect\subref{fig:3delay}, and \protect\subref{fig:loop} Schematic depictions of the real systems. \protect\subref{fig:delay cascade}, \protect\subref{fig:2delay cascade}, \protect\subref{fig:3delay cascade}, and \protect\subref{fig:loop cascade} The corresponding cascades of system copies, each for five intervals. \protect\subref{fig:delay results}, \protect\subref{fig:2delay results}, \protect\subref{fig:3delay results}, and \protect\subref{fig:loop results} Simulation results. In each figure, the excitation number in the system with delayed feedback is shown in red; the corresponding result without feedback is shown in black for comparison. The result for an undriven system (calculated using classical delay-differential equations~\cite{milonni_retardation_1974,dorner_laser-driven_2002,tufarelli_dynamics_2013}) is shown in blue. \protect\subref{fig:delay results} Results for the single loop, with drive $\Omega/\gamma=1/4$. \protect\subref{fig:2delay results}, \protect\subref{fig:3delay results} and \protect\subref{fig:loop results} Results for the two-loop, three-loop and infinite loop systems respectively, with $\Omega/\gamma=1$.\label{fig:1}}
\end{figure}

\begin{figure}
  \begin{minipage}[b]{.49\textwidth}
  \centering
  \subfloat[][\label{fig:equal delays}]{
    \includegraphics[page=13]{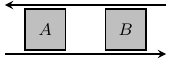}
  }
  \subfloat[][\label{fig:equal delays cascade}]{
    \includegraphics[page=14]{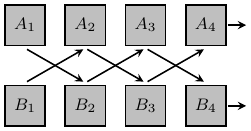}
  }\\
  \subfloat[][\label{fig:equal delays results}]{
    \includegraphics[page=15]{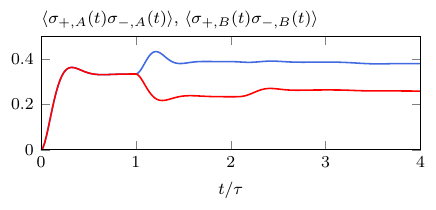}
  }
  \end{minipage}
  \begin{minipage}[b]{.49\textwidth}
  \subfloat[][\label{fig:different delays}]{
    \includegraphics[page=16]{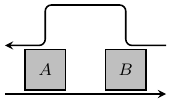}
  }
  \subfloat[][\label{fig:different delays cascade}]{
    \includegraphics[page=17]{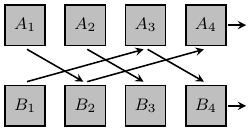}
  }\\
  \subfloat[][\label{fig:different delays results}]{
    \includegraphics[page=18]{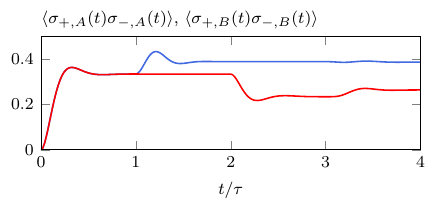}
  }
  \end{minipage}
  \caption{Cascaded systems with delayed backscatter. \protect\subref{fig:equal delays} and \protect\subref{fig:different delays} Schematic depictions of the real systems. \protect\subref{fig:equal delays cascade} and \protect\subref{fig:different delays cascade} Schematic depictions of the corresponding cascades of system copies, each for four intervals. \protect\subref{fig:equal delays results} and \protect\subref{fig:different delays results} Simulation results with parameters $\gamma\tau = 5$, $\phi = \pi/2$, $\Omega/\gamma=1$. The excitation number in subsystem $A$ is shown in red, and $B$ in blue. \protect\subref{fig:equal delays}--\protect\subref{fig:equal delays results} Equal delays in each direction. \protect\subref{fig:different delays}--\protect\subref{fig:different delays results} Different delays in each direction. Here the propagation delay from $A$ to $B$ is $\tau$, while the delay from $B$ to $A$ is $2\tau$.\label{fig:2}}
\end{figure}

We turn now to our examples. Figures~\ref{fig:1} and~\ref{fig:2} illustrate four categories of system, listed below. In each case, we display a schematic depiction of the system in question, as well as a sketch of the corresponding array of cascaded systems; finally, we display the results of simulations (performed using a program based on QuTiP~\cite{qutip1,qutip2}) with selected parameters.

While the dissipation kernel Eq.~\eqref{eq:35} is quite general and describes multiple reservoirs, for simplicity we will initially focus on a single system coupled to a single reservoir. Firstly, we reproduce for comparison the now well-understood case of a driven qubit emitting into a feedback loop with a discrete delay~\cite{grimsmo_time-delayed_2015}. The feedback loop is described by the dissipation kernel
\begin{equation}
  \label{eq:39}
  F(t) = 2 \gamma \delta(t) + \gamma \big[ e^{i \phi} \delta(t - \tau) + e^{-i \phi} \delta(t + \tau) \big] \,.
\end{equation}
where $\phi = \omega_0 \tau$. The internal system Hamiltonian, describing coherent driving with Rabi frequency $\Omega/2$, is $\Omega(\sigma_- + \sigma_+)$. This system is depicted schematically in Fig.~\subref*{fig:delay}, and the corresponding cascade of system copies is shown in Fig.~\subref*{fig:delay cascade}. Our simulation results are shown in Fig.~\subref*{fig:delay results}.

Secondly, we examine a qubit, driven as above, that instead couples to the reservoir at $N$ locations with equal spacing $\tau$. This is described by the coupling constant $\kappa_j = \sum_{n=0}^{N-1} \sqrt{\gamma/N} e^{i \omega_j n \tau}$; the corresponding dissipation kernel is
\begin{equation}
  \label{eq:40}
  F(t) = 2 \gamma \delta(t) + 2 \gamma \sum_{n=1}^{N-1} \left(1 - 
\frac{n}{N}\right) \big[ e^{i n \phi} \delta(t - n\tau) + e^{-i n\phi} \delta(t + n\tau) \big] \,.
\end{equation}
Note that this dissipation kernel can also be thought of as describing $N$ separate reservoirs, each with a single feedback loop: from the perspective of the system alone, the dynamics are identical. Figures~\subref*{fig:2delay results} and \subref*{fig:3delay results} show sample simulations with the dissipation kernel~\eqref{eq:40}, with finite $N$. In the limit $N \to \infty$, this dissipation kernel describes the situation depicted in Fig.~\subref*{fig:loop}, in which the system couples to a reservoir that loops back on itself without any irreversible dissipation. This can be thought of as a system coupled to a (very long) multi-mode cavity which acts as a one-dimensional waveguide that feeds anything that gets into the waveguide back to the system first after one round trip, then also after two round trips, three, and so on.

Thirdly, we turn attention to the case of more than one reservoir: we consider cascaded qubits with backscatter, that is a pair of driven qubits arranged such that the output from each subsystem drives the other subsystem with a propagation delay $\tau$ in both directions. This system has previously been examined by Pichler and Zoller~\cite{pichler_photonic_2016} using a different technique. The reservoir is characterised by the dissipation kernels:
\begin{eqnarray}
  F_{AA}(t) = F_{BB}(t) = 2 \gamma \delta(t) \,, \label{eq:41} \\
  F_{BA}(t) = \gamma \big[ e^{i \phi} \delta(t - \tau) + e^{-i \phi} \delta(t + \tau) \big] \,, \label{eq:42} \\
  F_{AB}(t) = \gamma \big[ e^{i \phi} \delta(t - \tau) + e^{-i \phi} \delta(t + \tau) \big] \,. \label{eq:43}
\end{eqnarray}
The internal system Hamiltonian is given by $H = \Omega(\sigma_{-,A} + \sigma_{+,A}) + \Omega(e^{i\phi} \sigma_{-,B} + e^{-i\phi} \sigma_{+,B})$.

Finally, we generalise the model of cascaded qubits with backscatter to consider delays that differ in each direction. More specifically, we consider the case where the delay from subsystem $B$ to $A$ is twice the delay from $A$ to $B$. The reservoir is characterised by the kernels~\eqref{eq:41} and~\eqref{eq:42}, along with
\begin{equation}
  \label{eq:44}
  F_{AB}(t) = \gamma \big[ e^{i \phi} \delta(t - 2\tau) + e^{-i \phi} \delta(t + 2\tau) \big] \,.
\end{equation}
We have here supposed that the phase advance in each direction is the same. The internal system Hamiltonian is as in the previous example.

These examples are presented only to illustrate the applicability of the derived algorithm. Of course, there is much more that could be said about the physics of any one of them, or variations on the driven qubit setup.

\section{Two-time correlation functions}
\label{sec:two-time-correlation}

The algorithm can be further developed to allow computation of multi-time correlation functions. We consider here two-time correlation functions, though the method is easily generalised. Denoting by $\rho_{S+E}(t)$ the combined state of the system and reservoir, we trivially find
\begin{equation}
  \label{eq:45}
  \expect{A(t_1) B(t_2) C(t_1)} = \tr_S[ B \varrho_{S;CA}(t_2, t_1) ] \,, \qquad t_2 \geq t_1 \,,
\end{equation}
where
\begin{equation}
  \label{eq:46}
  \varrho_{S;CA}(t_2, t_1) \equiv \tr_E[ U(t_2, t_1) C \rho_{S+E}(t_1) A U^\dagger(t_2, t_1) ] \,,
\end{equation}
and where we recall that $U(t, t_0)$ is the unitary generated by $H_S + H_{SE}(t)$, with interaction part given by Eq.~\eqref{eq:13}. Equations~\eqref{eq:45} and~\eqref{eq:46} are sometimes referred to as the quantum regression formula. Once again, we divide the integration time $t$ into intervals of length $\xi$. As one might intuitively expect, the operators $C$ and $A$ are applied in the $l_1$\textsuperscript{th} fictitious system copy at time $t_1'$, with $t_n' = t_n - (l_n - 1)\xi$ with $l_n = \lceil t_n/\xi \rceil$. That is to say, we modify Eq.~\eqref{eq:3} to read:
\begin{eqnarray}
  \label{eq:47}
  \fl e_{j_1 \cdots j_n;CA,l_1}(\xi; t_2', t_1') \equiv \nonumber \\
  \Phi^{(n-1)}(\xi, t_2')
  \cases{
    \Phi^{(n)}(t_2', t_1') C_{l_1} \left[\Phi^{(n)}(t_1', 0) e_{j_1 \cdots j_n}\right] A_{l_1} & $t_1' \leq t_2'$ \,, \\
    C_{l_1} \left[\Phi^{(n-1)}(t_2', t_1') \Phi^{(n)}(t_1', 0) e_{j_1 \cdots j_n}\right] A_{l_1} & $t_1' > t_2'$ \,.
  }
\end{eqnarray}
We similarly modify Eq.~\eqref{eq:2}:
\begin{equation}
  \label{eq:48}
  e_{j_1;CA,l_1}(t_2, t_1) = \tr_1 \cdots \tr_{n-1} \sum_{j_2 \cdots j_n} e_{j_1 \cdots j_n;CA,l_1}(\xi; t_2', t_1') [e_{j_2 \cdots j_n}^\dagger \otimes I] \,.
\end{equation}
Once again it is a simple exercise to show that Eqs.~\eqref{eq:47} and~\eqref{eq:48} reconstruct the operator~\eqref{eq:46}. These equations therefore represent a generalisation of the quantum regression formula to the case of open quantum systems with delayed coherent feedback.

\begin{figure}
  \centering
  \includegraphics[page=19]{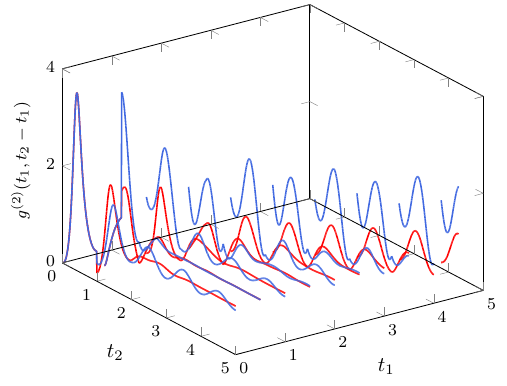}
  \caption{Second order photon correlation functions of the output field from a driven qubit emitting into a feedback loop with a single discrete delay, calculated using Eq.~\eqref{eq:49}, for $t_1 = 0, 0.5, \ldots, 4.5$. Simulations performed with delay given by $\gamma\tau = 1$ and drive $\Omega/\gamma = \pi$. Results for $\phi=0$ are shown in red, and $\phi=\pi$ in blue.}\label{fig:g2}
\end{figure}

We can use Eqs.~\eqref{eq:47} and~\eqref{eq:48} to calculate properties of the environment. As an example, consider the second-order photon correlation function
\begin{equation}
  \label{eq:49}
  g_\alpha^{(2)}(t_1, t_2 - t_1) = \frac{\expect{E_\alpha^\dagger(t_1) E_\alpha^\dagger(t_2) E_\alpha(t_2) E_\alpha(t_1)}}{\expect{E_\alpha^\dagger(t_1) E_\alpha(t_1)} \expect{E_\alpha^\dagger(t_2) E_\alpha(t_2)}} \,,
\end{equation}
$t_2 \geq t_1$, where $E_\alpha(t)$ is the field in the $\alpha$\textsuperscript{th} subenvironment outside the feedback loop, measured at the location of the output from the system. It is easy to show using Heisenberg picture methods~\cite{gardiner_input_1985,gardiner_input_1987,whalen_time-local_2016} that
\begin{equation}
  \label{eq:50}
  E_\alpha(t) = B_\alpha(t) - i \sum_\beta \int_0^t \D{t'} F_{\alpha\beta}(t - t') a_\beta(t') \,,
\end{equation}
which allows us to express $g_\alpha^{(2)}(t_1, t_2 - t_1)$ in terms of system operators. A sample calculation is presented in Fig.~\ref{fig:g2}, which shows the second order correlation function (in the transient regime) for the output field from a driven qubit emitting into a feedback reservoir with a single discrete delay, as described by the dissipation kernel~\eqref{eq:39}, for various different $t_1$.

\section{The cascade algorithm as a quantum teleportation protocol}
\label{sec:teleportation}

In this section we show that the mapping rule~\eqref{eq:2} may be interpreted as a quantum teleportation protocol~\cite{bennett_teleporting_1993}. For simplicity, we consider the evolution up to a time $\tau < t \leq 2 \tau$, for which we require two system copies in the fictitious chain of our algorithm; the generalisation to $n$ systems is straightforward.

Alice has two systems $S_0$ and $S_1$, while Bob has a single system $S_2$. All three systems are isomorphic. Systems $S_0$ and $S_2$ are prepared in the maximally-entangled state $D^{-1/2} \sum_{\mu} \ket{\mu} \otimes \ket{\mu}$, with $D$ the subsystem dimension, while $S_1$ is prepared in the initial system state $\rho(0)$. Thus, the initial state of the combined system can be written as
\begin{equation}
  \label{eq:51}
  \rho_{012} = \frac{1}{D} \sum_{\mu\nu} \ketbra{\mu}{\nu} \otimes \rho(0) \otimes \ketbra{\mu}{\nu} \,.
\end{equation}
An orthonormal basis of maximally entangled states is given by $\{\ket{\psi^{(pq)}}\}$ with
\begin{equation}
  \label{eq:52}
  \ket{\psi^{(pq)}} = \frac{1}{\sqrt{D}} \sum_{\mu} e^{2\pi i j p / D} \ket{\mu} \otimes \ket{\mu \oplus q} \,,
\end{equation}
where $\mu \oplus q \equiv (\mu + q) \bmod D$.

To recapitulate the conventional quantum teleportation protocol, suppose Alice performs a measurement on her systems $S_0$ and $S_1$ in this basis. The resulting state of Bob's system $S_2$, conditioned on Alice measuring the outcome $pq$, is
\begin{equation}
  \label{eq:53}
  \rho_2^{(pq)} = \sum_{\mu\nu} e^{2\pi i (\nu - \mu) p / D} \expect{\mu|\rho(0)|\nu} \ketbra{\mu \oplus q}{\nu \oplus q}  \,.
\end{equation}
If Bob then applies the unitary
\begin{equation}
  \label{eq:54}
  U^{(pq)} = \sum_\mu e^{2\pi i \mu p / D} \ketbra{\mu}{\mu \oplus q} \,,
\end{equation}
he is left with a copy of $\rho(0)$ in his system.

Now suppose Alice is also in possession of a time machine, so that she can send the outcome $pq$ of her measurement to Bob such that it reaches him before the measurement is performed. Bob applies the necessary unitary transformation to his system \emph{before} the experiment begins, resulting in the combined system state
\begin{equation}
  \label{eq:55}
  U_2^{(pq)} \rho_{012} U_2^{(pq)\dagger} = \frac{1}{D} \sum_{\mu} \ketbra{\mu}{\nu} \otimes \rho(0) \otimes U^{(pq)} \ketbra{\mu}{\nu} U^{(pq)\dagger} \,.
\end{equation}
To make contact with the cascade algorithm, the systems $S_1$ and $S_2$ -- belonging to Alice and Bob respectively -- are now acted upon by the map $\Phi^{(1)}(\tau, t') \Phi^{(2)}(t', 0)$, $t' = t - \tau$. The result is of course
\begin{eqnarray}
  \label{eq:56}
  \fl U_2^{(pq)} \rho_{012} U_2^{(pq)\dagger} = \frac{1}{D} \sum_{\mu\nu} \Bigl\{ \ketbra{\mu}{\nu} \otimes \Phi^{(1)}(\tau, t') \Phi^{(2)}(t', 0) \left[\rho(0) \otimes U^{(pq)} \ketbra{\mu}{\nu} U^{(pq)\dagger}\right] \Bigr\} \,.
\end{eqnarray}
Alice then performs the same measurement as before, sending the result back in time so that Bob can perform the correct unitary ahead of time. This procedure leaves Bob with the state
\begin{eqnarray}
  \label{eq:57}
  \rho_2^{(pq)} &= \frac{1}{D} \tr_1 \sum_{\mu\nu} e^{2\pi i (\nu - \mu) p / D} \Bigl\{ \Phi^{(1)}(\tau, t') \Phi^{(2)}(t', 0) \nonumber\\ &\quad \times \Bigl[\rho(0) \otimes U^{(pq)} \ketbra{\mu \oplus q}{\nu \oplus q} U^{(pq)\dagger}\Bigr] \Bigl[ \ketbra{\nu}{\mu} \otimes I \Bigr]\Bigr\} \nonumber \\
                &= \tr_1 \sum_{\mu\nu} \Phi^{(1)}(\tau, t') \Phi^{(2)}(t', 0) [\rho(0) \otimes \ketbra{\mu}{\nu} ] [\ketbra{\nu}{\mu} \otimes I] \,.
\end{eqnarray}
If we now make the same notational change as in Sec.~\ref{sec:closed-decomposition}, with a single index standing in for the pair $(\mu,\nu)$ and $e_{j}$ standing in for $\left(\ketbra{\mu}{\nu}\right)$, Eq.~\eqref{eq:57} becomes
\begin{equation}
  \label{eq:58}
  \rho_2^{(pq)} = \tr_1 \sum_{j} \Phi^{(1)}(\tau, t') \Phi^{(2)}(t', 0) [\rho(0) \otimes e_j] [e_j^\dagger \otimes I] \,,
\end{equation}
which is equivalent to Eqs.~\eqref{eq:1},~\eqref{eq:2} and~\eqref{eq:3} together in the case $n=2$. The need for time travel is of course unphysical, but the result can be reproduced in a probabilistic fashion: Bob applies no unitary (or, equivalently, $U^{(00)} = I$) and the protocol succeeds whenever Alice obtains the outcome $pq = 00$.

\section{Conclusion and outlook}
\label{sec:conclusion}

We have derived a generalisation of a technique, derived previously by one of us, for simulating feedback in open quantum systems. Our derivation uses only elementary methods and is based on decomposing the time evolution of a general open quantum system into intervals represented as separate system copies---that is to say, this decomposition is not limited to systems exhibiting delayed coherent feedback. The resulting simulation method admits multiple subsystems with multiple delays in cases where those delays are commensurable. We used our generalised method to simulate systems with multiple delays, including cascaded systems with delayed backscatter.

In addition, we presented a generalisation of the quantum regression formula that applies to systems with delayed feedback, and demonstrated how to use this formula to compute two-time correlation functions of the system and output field properties. Finally, we showed that delayed coherent feedback can be simulated through either an exotic quantum teleportation protocol requiring time travel, or through a probabilistic teleportation protocol.

We conclude with some general remarks on the relation between the techniques presented above and non-Markovian open quantum systems in general. Consider, for example, a single open quantum system that interacts with a feedback reservoir with a generic memory kernel $f(t)$. This memory kernel may be approximated by requiring that integrals over it become left Riemann sums:
\begin{equation}
  \label{eq:59}
  f(t) \approx h \sum_{j=0}^\infty \left[ f(hj) \delta(t - hj) + f(-hj) \delta(t + hj) \right] \,,
\end{equation}
for some chosen $h$. Note that Eq.~\eqref{eq:59} takes the same form as Eq.~\eqref{eq:35} provided $f(t)$ is Hermitian. The error in this approximation is $O(h^2)$, and the exact memory kernel is of course recovered in the limit $h \to 0$. As such, provided we have the computational resources to consider sufficiently small $h$, any memory kernel -- even a continuous one -- may be approximated as a series of discrete delayed feedback loops. Because of this, continuous coherent feedback can be viewed as an infinite chain of cascaded system copies, subject once again to the inter-system mapping formula~\eqref{eq:2}. The method owes its conceptual generality to the ability of this mapping rule to insert (or teleport, as shown in Sec.~\ref{sec:teleportation}) a history into the system's evolution after the fact.

\ack{
  We would like to thank the anonymous referee for pointing out an error in an earlier version of the manuscript. This work was supported by the Marsden Fund of the Royal Society of New Zealand.%
}

\appendix

\section{Derivation of Eq.~\eqref{eq:26}}
\label{sec:deriv-divisible}

We briefly discuss here the reason why Eq.~\eqref{eq:26} is true whenever $\Phi^{(m)}$ is divisible, for all $m$. We can define an auxiliary map $\Phi_{t'}$ that satisfies
\begin{equation}
  \label{eq:60}
  \Phi_{t'}(t_1, t_0) \chi \equiv \tr_E \{\tilde{U}_{t'}(t_1, t_0) [ \rho_{\tilde{E}} \otimes \chi] \tilde{U}_{t'}^\dagger(t_1, t_0) \} \,,
\end{equation}
where
\begin{equation}
  \label{eq:61}
  \tilde{U}_{t'}(t_1, t_0) = 1_{t_0 \leq t_1 < t'} \tilde{U}^{(n)}(t_1, t_0) + 1_{t' \leq t_0} \tilde{U}^{(n-1)}(t_1, t_0) \,.
\end{equation}
As such, we have
\begin{equation}
  \label{eq:62}
  \Phi_{t'}(t_1, t_0) = 1_{t_0 \leq t_1 < t'} \Phi^{(n)}(t_1, t_0) + 1_{t' \leq t_0} \Phi^{(n-1)}(t_1, t_0)
\end{equation}
A sum of divisible maps, with positive real coefficients, is also divisible~\cite{rivas_quantum_2014,kossakowski_quantum_1972,gorini_completely_1976,lindblad_generators_1976}. Because the indicator functions $1_{\cdots}$ take the non-negative values $0$ or $1$, if both $\Phi^{(n)}$ and $\Phi^{(n-1)}$ are divisible -- as we have assumed -- then $\Phi_{t'}$ must be divisible as well. As such, we may write
\begin{eqnarray}
  \label{eq:63}
  \fl \Phi_{t'}(\xi, t') \Phi_{t'}(t', 0) \chi = \Phi_{t'}(\xi, 0) \chi = \tr_E \{\tilde{U}_{t'}(\xi, t') \tilde{U}_{t'}(t', 0) [ \rho_{\tilde{E}} \otimes \chi] \tilde{U}_{t'}^\dagger(t', 0) \tilde{U}_{t'}^\dagger(\xi, t')\} \,. \nonumber\\
\end{eqnarray}
Eq.~\eqref{eq:26} then follows immediately from the definitions~\eqref{eq:61} and~\eqref{eq:62}.

\section{Derivation of Eq.~\eqref{eq:33}}
\label{sec:derivation-eff-kernel}

Our aim is to show that the environment operator $\tilde{B}_{\alpha;m}(t')$ appearing in Eq.~\eqref{eq:22} may be chosen such that Eq.~\eqref{eq:32} agrees with Eq.~\eqref{eq:21}. The unitary~\eqref{eq:17} that appears in Eq.~\eqref{eq:21} may be re-written as
\begin{eqnarray}
  \label{eq:64}
  \fl U'(t_1, t_0) \equiv T \exp\Biggl( -i \int_{t_0}^{t_1} \D{s} \sum_m \Biggl\{1_{(m-1) \xi \leq s < m \xi} H_{S;m} \nonumber\\ \hspace*{3.5cm} + \sum_\alpha \left[ B_{\alpha;m}'(s) \otimes a_{\alpha;m}^\dagger + B_{\alpha;m}^{\prime\dagger}(s) \otimes a_{\alpha;m} \right]\Biggr\} \Biggr) \,,
\end{eqnarray}
where we have defined
\begin{equation}
  \label{eq:65}
  B_{\alpha;m}'(s) \equiv 1_{(m-1) \xi \leq s < m \xi} B_\alpha(s) \,.
\end{equation}
Equation~\eqref{eq:32} may be written in a form analogous to Eq.~\eqref{eq:21}:
\begin{equation}
  \label{eq:66}
  \fl e_{S;j_1 \cdots j_n}(\xi; t') = \tr_E \left\{ \tilde{U}'(t' + (n-1)\xi, 0) \left[ \rho_{\tilde{E}} \otimes e_{S;j_1 \cdots j_n} \right] \tilde{U}^{\prime\dagger}(t' + (n-1)\xi, 0) \right\} \,,
\end{equation}
where we have defined
\begin{eqnarray}
  \label{eq:67}
  \fl \hspace*{1cm} \tilde{U}'(t_1, t_0) &\equiv T \exp\left( -i \int_{t_0}^{t_1} \D{s} \sum_m 1_{(m-1) \xi \leq s < m \xi} \tilde{H}_m(s - (m + 1)\xi) \right) \nonumber \\
  &\equiv T \exp\Biggl( -i \int_{t_0}^{t_1} \D{s} \sum_m \Biggl\{1_{(m-1) \xi \leq s < m \xi} H_{S;m} \nonumber\\ &\hspace*{3.5cm} + \sum_\alpha \left[ \tilde{B}_{\alpha;m}'(s) \otimes a_{\alpha;m}^\dagger + \tilde{B}_{\alpha;m}^{\prime\dagger}(s) \otimes a_{\alpha;m} \right]\Biggr\} \Biggr) \,,
\end{eqnarray}
with
\begin{equation}
  \label{eq:68}
  \tilde{B}_{\alpha;m}'(s) \equiv 1_{(m-1) \xi \leq s < m \xi} \tilde{B}_{\alpha;m}(s - (m + 1)\xi) \,.
\end{equation}

Observe that the unitaries~\eqref{eq:64} and~\eqref{eq:67} differ only in terms of the environment operators~\eqref{eq:65} and~\eqref{eq:68}. If we restrict attention to vacuum reservoirs (as we do in Sec.~\ref{sec:effective-dissipation-kernel}), the right-hand sides of Eqs.~\eqref{eq:21} and~\eqref{eq:66} are equal if the two interactions have equivalent dissipation kernels. To be more specific, we need
\begin{equation}
  \label{eq:69}
  \commutator{\tilde{B}_{\alpha;m}'(t_2)}{\tilde{B}_{\beta;m'}^{\prime\dagger}(t_1)} = \commutator{B_{\alpha;m}'(t_2)}{B_{\beta;m'}^{\prime\dagger}(t_1)} \,,
\end{equation}
for all $t_2 \geq t_1$, which is satisfied provided
\begin{equation}
  \label{eq:70}
  \commutator{\tilde{B}_{\alpha;m}(t_2 - (m + 1)\xi)}{\tilde{B}_{\beta;m'}^\dagger(t_1 - (m' + 1)\xi)} = \commutator{B_\alpha(t_2)}{B_\beta^\dagger(t_1)} \,,
\end{equation}
or equivalently
\begin{equation}
  \label{eq:71}
  \tilde{F}_{\alpha\beta;mm'}(t_2 - t_1) = F_{\alpha\beta}(t_2 - t_1 + (m - m') \xi) \,,
\end{equation}
for all $t_2 \geq t_1$ and $m \geq m'$. Combining Eq.~\eqref{eq:71} with Eq.~\eqref{eq:29} immediately gives Eq.~\eqref{eq:33}.

\section*{References}
\bibliography{references.bib}

\end{document}